\documentclass{article}
\usepackage{amsmath,amsthm}
\usepackage{amssymb,latexsym}
\usepackage[mathscr]{eucal}
\usepackage{setspace}
\usepackage{graphics}
\usepackage{array}

\newcommand{\beq}{\begin{equation}}
\newcommand{\eq}{\end{equation}}
\newcommand{\pa}{\partial}

\newcommand{\bfv}{{\bf v}}

\newcommand{\bfo}{{\bf \omega}}
\newcommand{\rp}{\right)}
\newcommand{\lp}{\left(}
\newcommand{\rb}{\right]}
\newcommand{\lb}{\left[}
\newcommand{\orho}{\lp\frac{\omega}{\rho}\rp}
\newcommand{\ul}{_L}
\newcommand{\vl}{\int_{V\ul}}

\usepackage{graphicx}
\graphicspath{{converted_graphics/}}
\begin{document}

\title{\bf Beltrami States for Compressible Barotropic Flows}         
\author{B.K. Shivamoggi\footnote{Permanent Address: University of Central Florida, Orlando, FL 32816-1364} and G.J.F. van Heijst\\
Technische Universiteit Eindhoven\\
NL-5600 MB Eindhoven\\
The Netherlands}        
\date{\today}          
\maketitle

\large{\bf Abstract}

Beltrami states for compressible barotropic flows are deduced by minimizing the total kinetic energy while keeping the total helicity constant. A Hamiltonian basis for these Beltrami states is also sketched.

\pagebreak

\section{Introduction}       

It is well known (Moffatt \cite{Mof1}) that a significant class of exact solutions of the fluid dynamics equations emerges under the Beltrami condition - the local vorticity is proportional to the stream function. These Beltrami solutions are also known to correlate well with real fluid behavior - ex. the Larichev-Reznik \cite{Lar} nonlinear dipole-vortex localized structure. The purpose of this paper is to give the Beltrami states for compressible flows. Immediate difficulties arise for the latter case, because the pressure field is now determined by thermodynamics and hence plays a dynamical role (and is no longer a passive variable as in incompressible flows where it simply adjusts instantaneously so as to keep the velocity field solenoidal and satisfy the given boundary conditions). Some of these difficulties are mitigated by assuming the fluid to be barotropic, i.e., the pressure is a single-valued function of the mass density - without this assumption, the compressible flow equations have to be closed by adding an equation of state and an equation for the evolution of internal energy.

\section{Beltrami States for Compressible Flows}

The equations governing compressible flows are (in usual notations) - 

\beq
\frac{D{\bf v}}{Dt}=-\frac{1}{\rho}\nabla p
\eq

\beq
\frac{D{\rho}}{Dt} +\rho\nabla \cdot {\bf v}=0
\eq

\noindent
from which, we have

\beq
\frac{\pa\bfo}{\pa t} = \nabla \times (\bfv\times\bfo) -\nabla \times \lp\frac{1}{\rho}\nabla p\rp.
\eq

Assuming barotropic condition - 

\beq
\nabla P \equiv \frac{1}{\rho}\nabla p \ or \ P \equiv \int\frac{dp}{\rho}.
\eq

\noindent
equation (3) becomes

\beq
\frac{\pa \omega}{\pa t} = \nabla \times (\bfv \times \bfo).
\eq

The Beltrami state is therefore given by

\beq
\bfo = a \bfv.
\eq

\noindent
$a$ being an arbitrary function of space and time.

In order to see if this state has a variational characterization, i.e. if it is a minimizer of energy on an isohelicity surface, first note that equation (5), on using equation (2), can be rewritten as 

\beq
\frac{D}{D t}\orho=\orho\cdot\nabla\bfv.
\eq

Using equation (7), we have

\beq
\frac{D}{D t}\lp \frac{\bfv\cdot\bfo}{\rho}\rp=\orho\cdot\nabla\lp\frac{1}{2}\bfv^2 - P\rp.
\eq

Let $S\ul$ be a surface enclosing a volume $V\ul$ and moving with the fluid; the total helicity is given by 

\beq
H_e \equiv \int_{V\ul} \bfv\cdot\bfo\  dV.
\eq

On noting the mass-conservation condition for a fluid element -  

\beq
\frac{D}{D t}(\rho\ dV)=0
\eq

\noindent
we have (Moffatt \cite{Mof2}),

\beq
\frac{d H_e}{d t}  = \int_{V\ul} \frac{D}{D t} \lp \frac{\bfv\cdot\bfo}{\rho}\rp \rho\ dV\notag
\eq

\medskip		  
                   
\beq
= \int_{V\ul} (\bfo\cdot\nabla)\lp\frac{1}{2}\bfv^2 -P\rp dV \notag
\eq

\medskip

\beq
= \int_{S\ul} ({\bf {\hat n}}\cdot\bfo)\lp\frac{1}{2}\bfv^2 -P\rp dS.
\eq

\medskip
Assuming ${\bf \hat n}\cdot\bfo=0$ on $S\ul$, we obtain 

\beq
H_e=const.
\eq

Consider now the minimizer of the energy - 

\beq
E \equiv \int_{V\ul}\frac{1}{2}\rho\bfv^2 dV
\eq

\noindent
on an isohelicity surface - 

\beq
H_e = \int_{V\ul} \bfv\cdot\bfo\ dV = const.
\eq

It is given by 

\begin{subequations}

\beq
\delta \vl \lb\frac{1}{2}\rho\bfv^2 + \lambda(\bfv\cdot\bfo)\rb dV = 0
\eq

\noindent
or

\beq
\vl \lb\delta \lp\frac{\bfv^2}{2}\rp + \lambda\ \delta\lp\frac{\bfv\cdot\bfo}{\rho}\rp\rb(\rho\ dV)=0
\eq

\noindent
or

\beq
\vl\lb\bfv\cdot\delta\bfv+\lambda \left\{\lp\frac{\bfo}{\rho}\rp\cdot\delta\bfv+\bfv\cdot\delta\orho\right\}\rb(\rho\ dV) = 0.
\eq

\end{subequations}

\medskip
\noindent
which shows that a sufficient condition for the minimizer is that

\beq
\bfo=b \rho \bfv
\eq

\noindent
This is just the Beltrami state (6), with $ a = b\rho $!

Using (16), equation (1) gives, for the Beltrami state,

\beq
P+\frac{1}{2}\bfv^2=const
\eq

\noindent
everywhere, which, for the incompressible case, reduces to 

\beq
\frac{p}{\rho}+\frac{1}{2}\bfv^2 = const.
\eq

\section{Hamiltonian Basis for Beltrami States}

\indent
The Hamiltonian for the system of equations (1) and (2) is 

\beq
H=\frac{1}{2}\int_V \psi \cdot \bfo\ dV
\eq

\noindent
where,

\beq
\rho\bfv = \nabla \times \psi
\eq

\noindent
with the gauge condition - 

\beq
\nabla \cdot \psi = 0.
\eq

(20) implies $\pa\rho/\pa t= 0$ - but this suffices if the goal is to get a grip over the final Beltrami state rather than follow the actual dynamics of the Beltramization process. One might want to explore the dynamics more closely, but this is generally a more difficult problem.

Let us choose $\bfo$ to be the canonical variable and take the skew-symmetric operator to be 

\beq
J\equiv -\nabla\times\lb\orho\times\lp\nabla\times(\cdot)\rp\rb.
\eq

The Hamilton equation is then 

\beq
\frac{\pa\bfo}{\pa t} = J\frac{\delta H}{\delta \bfo}  = -\nabla \times \lb\orho\times(\nabla \times \psi)\rb\notag
\eq

\beq							
= -\nabla\times \lb\orho\times\rho \bfv\rb\notag
\eq
						
\beq
= \nabla \times (\bfv\times\bfo)
\eq

\noindent
as required!

The Casimic invariants for this system are the solutions of

\beq
J\frac{\delta \mathit{C}}{\delta \bfo} = -\nabla \times\lb\orho\times\lp\nabla\times\frac{\delta C}{\delta \bfo}\rp\rb = {\bf 0}
\eq

\noindent
from which,

\beq
\frac{\delta C}{\delta \bfo}=\bfv
\eq

\noindent
so,

\beq
C = \int_V \bfv\cdot \bfo\ dV
\eq

\noindent
which is just the total helicity (9)!

The Beltrami state is the minimizer of H keeping $C$ constant, and is given by

\begin{subequations}
\beq
\frac{\delta H}{\delta \bfo} = \lambda \frac{\delta C}{\delta \bfo}
\eq

\noindent
or

\beq
\psi = \lambda \bfv
\eq

\noindent
or

\beq
\rho \bfv = \lambda \bfo
\eq

\end{subequations}

\noindent
as before!

\section{Discussion}

\indent
In view of the emergence of a significant class of exact solutions of equations of fluid flows under the Beltrami condition and their correlation to real fluid behavior, one may wonder whether fluids have an intrinsic tendency towards Beltramization.

Though we do not have good understanding of this aspect, it is known (Moffatt \cite{Mof1}) that Beltramization provides the means via which the underlying system can accomplish ergodicity of the streamlines. This follows by noting that Beltramization corresponds to relaxation of the constraint

\beq
\bfv \cdot \nabla \lp P + \frac{1}{2}\bfv^2\rp=0
\eq

\noindent
in steady compressible flows governed by

\beq
\bfv \times \bfo =\nabla \lp P + \frac{1}{2} \bfv^2\rp
\eq

\noindent
so that the streamlines are no longer confined to the surfaces given by 

\beq
P + \frac{1}{2} \bfv^2 = const
\eq

\noindent
and become ergodic.

\pagebreak

\section{Acknowledgements}

Our thanks are due to Professor Leon Kamp for helpful discussions.

\end{document}